\documentclass[prl,twocolumn,superscriptaddress,longbibliography]{revtex4-1}
\usepackage{graphicx}
\usepackage{latexsym}
\usepackage{amssymb}
\usepackage{amsmath}
\usepackage{amsfonts}
\usepackage{upgreek}
\usepackage{bm}
\usepackage{multirow}
\usepackage{color}
\usepackage{hyperref}
\hypersetup{
colorlinks = true,
linkcolor = [rgb]{0.70,0.13,0.13},
citecolor = [rgb]{0.13,0.55,0.13},
urlcolor = [rgb]{0.25, 0.41, 0.88}}

\newcommand{\dsR}{\mathbb{R}}

\newcommand{\scL}{\mathcal{L}}

\renewcommand{\Re}{\operatorname{Re}}
\renewcommand{\Im}{\operatorname{Im}}

\newcommand{\mat}[1]{\left[\begin{matrix}#1\end{matrix}\right]}

\newcommand{\eq}[1]{\begin{equation}#1\end{equation}}
\newcommand{\eqs}[1]{\begin{equation}\begin{split}#1\end{split}\end{equation}}
\newcommand{\eqnref}[1]{Eq.\,\eqref{#1}}
\newcommand{\figref}[1]{Fig.\,\ref{#1}}

\begin{document}

\title{Emergent Quantum Mechanics in an Introspective Machine Learning Architecture}
\author{Ce Wang}
\author{Hui Zhai}
\email{hzhai@tsinghua.edu.cn}
\affiliation{Institute for Advanced Study, Tsinghua University, Beijing 100084, China}
\author{Yi-Zhuang You}
\email{yzyou@physics.ucsd.edu}
\affiliation{Department of Physics, University of California, San Diego, CA 92093, USA}

\date{\today}
\begin{abstract}
Can physical concepts and laws emerge in a neural network as it learns to predict the observation data of physical systems? As a benchmark and a proof-of-principle study of this possibility, here we show an introspective learning architecture that can automatically develop the concept of the quantum wave function and discover the Schr\"odinger equation from simulated experimental data of the potential-to-density mappings of a quantum particle. This introspective learning architecture contains a machine translator to perform the potential to density mapping, and a knowledge distiller auto-encoder to extract the essential information and its update law from the hidden states of the translator, which turns out to be the quantum wave function and the Schr\"odinger equation. We envision that our introspective learning architecture can enable machine learning to discover new physics in the future.  
\end{abstract}
\maketitle

\begin{figure}[t]
\begin{center}
\includegraphics[width=0.95\columnwidth]{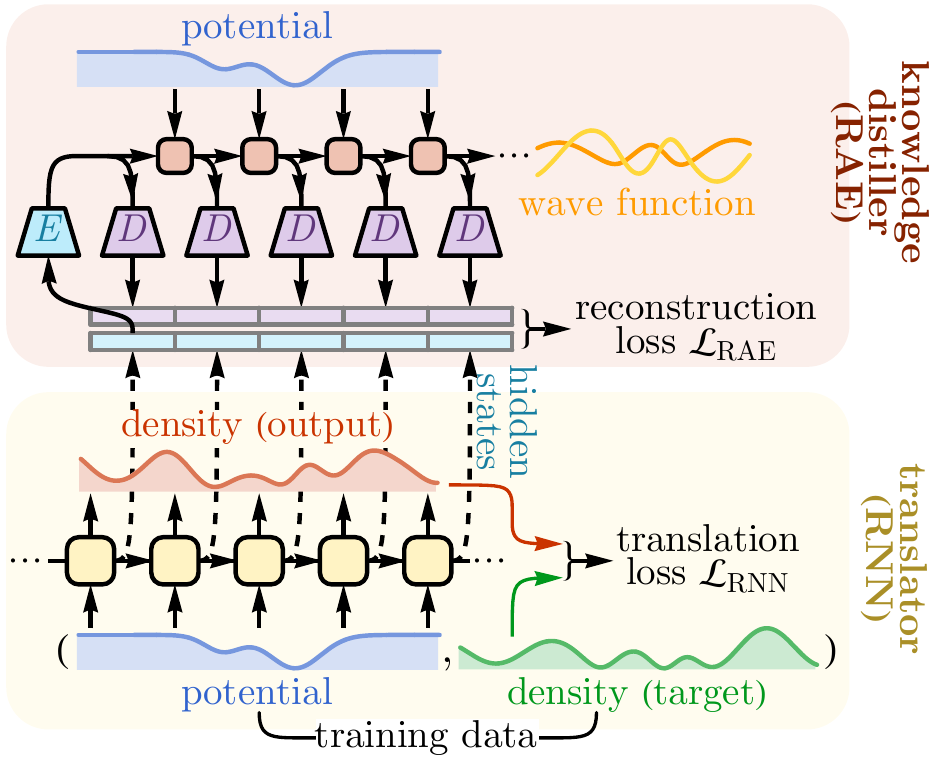}
\caption{The architecture of an introspective recurrent neural network, called ``the Schr\"odinger machine". It contains a translator (lower panel) and a knowledge distiller (upper panel). The translator is implemented as a recurrent neural network to perform the task of the potential-to-density mapping. The knowledge distiller compresses the hidden states generated by the translator using a recurrent auto-encoder and extracts the most essential variables in the hidden states together with its update rule. }
\label{fig:scheme}
\end{center}
\end{figure}

The ongoing third wave of artificial intelligence has made great achievements in employing neural-network-based machine learning for industry and social applications. Inspired by this great success, machine learning algorithms have also been rapidly applied to various directions of physics research, ranging from high-energy and string theory to condensed matter, atomic, molecular and optical physics.\cite{Carifio:2017Ma,Koch-Janusz:2018Mu,You:2018Ma,Hashimoto:2018DL,Torlai:2016Le,Wang:2016Di,Carrasquilla:2017Ma,van-Nieuwenburg:2017Le,Zhang:2017Qu,Wang:2017Ma,Wang:2018Ma,Zhang:2018MT} While there has been many successful examples of machine assisted physics research, it remains an ambitious goal to explore the potential of machine learning in unsupervised discovery of concepts and laws of physics from observation data.\cite{Iten:2018Di, Wu2018To} A major challenge is to understand how the machine ``thinks'', or what approaches have been developed inside its mind. This typically requires us to open up the black box of the neural network and to identify the most relevant emergent features in the neural activity. Can the analysis of the neural activity also be automated by the machine itself? Can knowledge emerges as the machine examines its own information flow introspectively? To demonstrate these possibilities, here we report an introspective learning architecture, as illustrated in \figref{fig:scheme}, that allows the machine to distill the knowledge about quantum mechanics from the observation of the density distributions of a quantum particle in different shapes of potentials.

As a proof-of-concept study, we consider a single quantum particle moving in a one-dimensional space with certain potential. Suppose we can measure the particle density for each given potential, we supply the machine with the potential profile as the input and the density profile as the target, and challenge the machine to discover the underlying rule governing the potential-to-density mapping. 
We discretize the potential $V(x)$ and density profiles $\rho(x)$ along the one-dimensional space and treat them as sequences of real numbers: $V_i=V(x_i)$ and $\rho_i=\rho(x_i)$, where $x_i=ai$ are the discrete coordinates for $i=0,1,2,\cdots$, which are evenly distributed along the one-dimensional space with a fixed separation $a=0.1$. We assume that the potential is always measured with respect to the energy of the particle, such that the particle energy is effectly fixed at zero. We will only consider the case of $V_i<0$, such that the particle remains in extended states. 

\begin{figure}[t]
\begin{center}
\includegraphics[width=0.82\columnwidth]{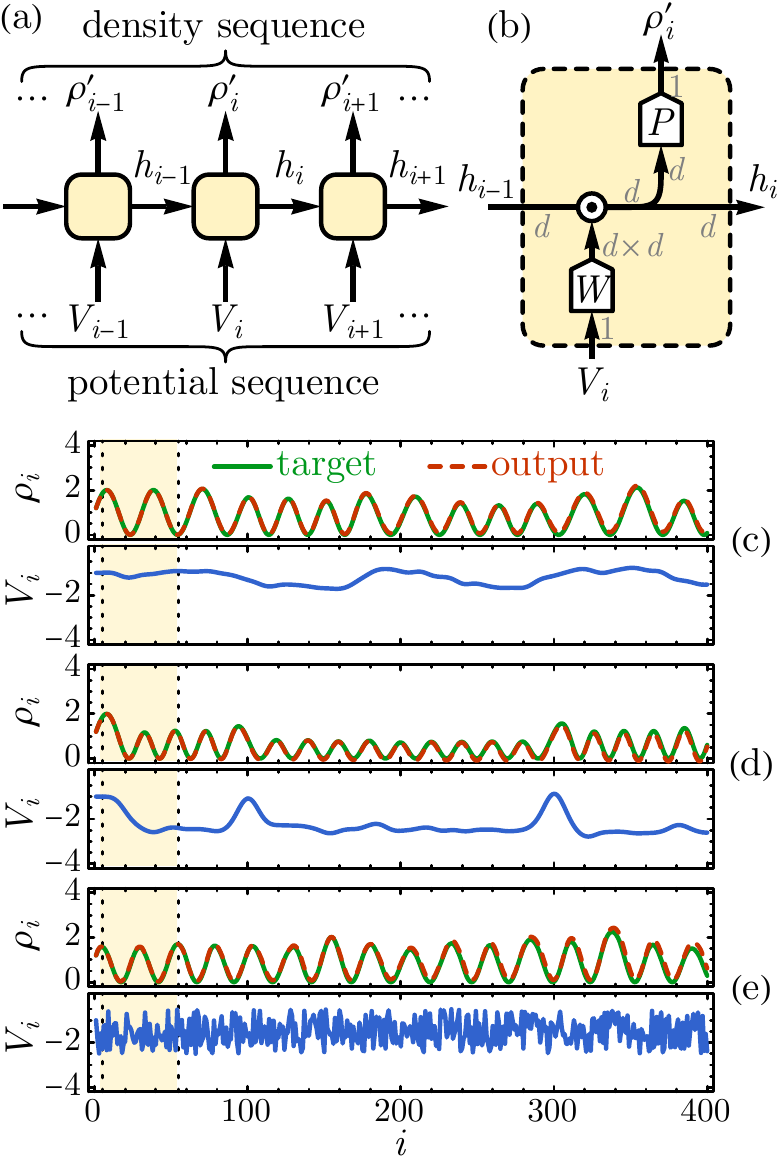}
\caption{Architecture of the translator RNN for the potential-to-density mapping. (a) is the global structure and (b) is the network structure within each block. Arrows indicate the direction of information flow. The tensor dimensions are marked out in gray. $W$ and $P$ can be generic functions, although they are modeled by the Taylor expansions in our implementation. The symbol $\odot$ denotes matrix-vector multiplication. (c-e) Typical samples of the RNN output density profiles in comparison with the target density profiles for (c) a shallow and smooth potential, (d) a deep but smooth potential, and (e) a shallow but rough potential. The model is only trained on a small window indicated by the yellow shaded region. The trained RNN can perform the potential-to-density mapping over a much larger range.}
\label{fig:RNN}
\end{center}
\end{figure}

By treating both the potential and density profiles as sequential data, the potential-to-density problem belongs to a broader class of sequence-to-sequence mapping,\cite{Kalchbrenner:2013Re,Sutskever:2014Se,Cho:2014Le,Bahdanau:2014Ne} which can be handled by the recurrent neural network (RNN).\cite{Goodfellow:2016De} The RNN has been widely used in natural language processing to translate sequences of words from the source language to the target language.\cite{Neubig2017Ne} We apply the RNN architecture to perform the potential-to-density mapping as a translation task.  In each step, the RNN takes an input $V_i$ from the source sequence, modifies its internal hidden state $h_i$ accordingly, and generates the output $\rho'_i$ based on the  hidden state, as illustrated in \figref{fig:RNN}(a). We adopt the following update equations
\eq{\label{eq:RNNupdate}
h_{i}=W(V_i)\cdot h_{i-1},\quad\rho'_i=P(h_i).
}
where both the input $V_i\in\dsR $ and the output $\rho'_i\in\dsR$ are scalars and the hidden state $h_i\in\dsR^d$ is a $d$-dimensional vector. The hidden state $h_i$ is updated by an input-dependent linear transformation, represented by a $d\times d$ matrix $W(V_i)\in\dsR^{d\times d}$ multiplied to the vector $h_i$. The output $\rho'_i$ is generated from the hidden state by a projection map $P(h_i)$. The data flow is graphically represented in \figref{fig:RNN}(b). The output sequence $\rho'_i$ is then compared with the target sequence $\rho_i$ over a window of steps to evaluate the loss function
\eq{\label{eq:loss_RNN}
\scL_\text{RNN}=\sum_{i\in\text{window}}(\rho'_i-\rho_i)^2.
}
How the RNN updates its hidden state and generates output is determined by the functions $W$ and $P$. In general, $W$ and $P$ could be non-linear functions modeled by feedforward neural networks for instance. However, for our problem, we find it sufficient to model $W$ by a Taylor expansion (to the $n_W$th order in $V_i$) and $P$ by a linear projection,
\eq{\label{eq:WP}
W(V_i)=\sum_{n=0}^{n_W} W^{(n)} V_i^n,\quad P(h_i)=p^\intercal\cdot h_i,
}
where $W^{(n)}$ is the $n$th order Taylor expansion coefficient matrix (each of the dimension $d\times d$) and $p$ is a $d$-dimensional vector. The elements in $W^{(n)}$ and $p$ are model parameters to be trained to minimize the loss function $\scL_\text{RNN}$. The training dataset contains pairs of potential and density sequences that serve as parallel corpora to train the RNN translator. They are currently obtained from numerical simulation \footnote{see Supplementary Material for details about data acquisition.}, but can be collected from experiments in future applications, from instance, the quantum gas microscope can detect density of ultracold atoms nearly in their ground state in-situ in the presence of different kind of potentials generated by optical speckles.\cite{Lye2005Bo} After minimizing the translator loss $\scL_\text{RNN}$, the RNN can predict the density profile based on the potential profile \footnote{see Supplementary Material for training method.}.

We build the RNN with the Taylor expansion order $n_W=2$ and the hidden state dimension up to $d=6$. We observe that the loss $\scL_\text{RNN}$ will drop significantly as long as $d\geq 3$ \footnote{see Supplementary Material for detailed analysis.}. Using the RNN model for the one-dimensional potential-to-density mapping is physically grounded because it respects the translational symmetry of the physical law that governs this mapping. As a result, an immediate advantage of the RNN is to gain spatial scalability, that is, what has been learned over a small system can be readily generalized and applied to larger systems. For instance, as shown in \figref{fig:RNN}(c-e), the RNN is trained over a small window from $i=5$ to $i=55$ (the initial 5 outputs are excluded to reduce the sensitivity to initial conditions). After training, the RNN can perform the potential-to-density mapping for a much larger system, from $i=0$ to $i=400$. \figref{fig:RNN}(c-e) shows that the RNN output matches nicely with the target density profile (with about 10\% relative error) on the test dataset for different classes of potential profiles, either shallow or deep, and either smooth or rough. This result demonstrates the prediction power of the RNN model.

By learning to perform the potential-to-density mapping, the RNN translator must have developed some intuitions about the underlying physics. Historically, advances in physics are often marked by formulating physical phenomena in term of differential equations, such as Newton's law of motion, Maxwell's equation of electromagnetism, and the Schr\"odinger equation of quantum mechanics. The RNN provides a universal representation of recurrent equations as discretized versions of the differential equations, and therefore the update rules of its hidden state can be interpreted as machine's understanding of the physical laws.\cite{Ma:2018Mo, Banchi:2018Mo} As the RNN performs the translation, it generates a sequence of hidden states containing the essential variables governing the physics of potential-to-density mapping, mixed with other redundant or irrelevant information. To extract the knowledge from these hidden state data, we design a higher-level machine, called the knowledge distiller, to learn from the neural activity (the hidden state sequence) of the lower-level translator. It works on the RNN hidden states to compress the information and to extract the underlying rule. The auto-encoder architecture is widely used for information compression.\cite{Bengio2013Re, Kingma2013Au} Here we incorporate the auto-encoder in another recurrent neural network structure as a recurrent auto-encoder (RAE), because we not only need to find out the essential variables in the hidden states but also need to determine the update rules of these essential variables.

\begin{figure}[t]
\begin{center}
\includegraphics[width=0.9\columnwidth]{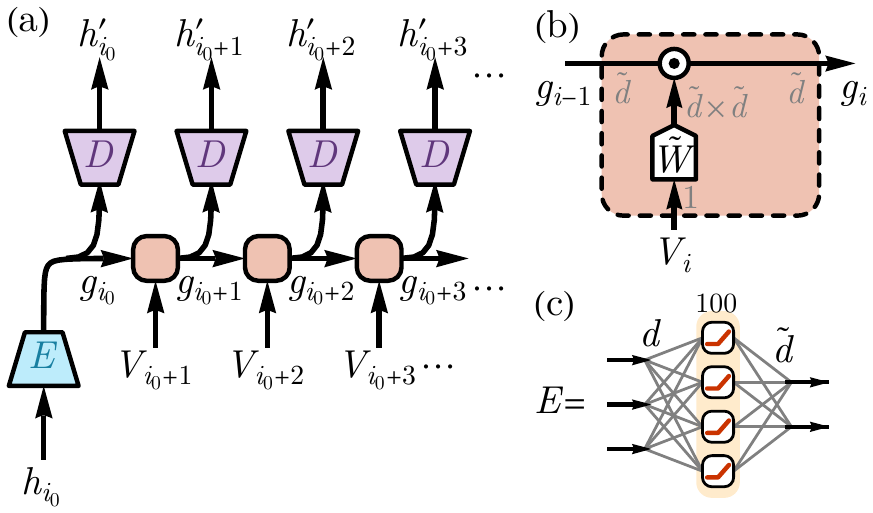}
\caption{Architecture of the recurrent auto-encoder. (a) The global structure. (b) The network structure within each recurrent block. (c) The feedforward network of the encoder $E$. Arrows indicate the direction of information flow. Tensor dimensions are marked out in gray. In (b), $\tilde{W}$ can be a generic function. The symbol $\odot$ denotes matrix-vector multiplication. In (c), we use one hidden layer of 100 dimension, with the ReLU activation. The decoder $D$ has a similar feedforward network in a revered structure as (c).}
\label{fig:RAE}
\end{center}
\end{figure}

The architecture of the RAE knowledge distiller is illustrated in \figref{fig:RAE}. The RAE distiller first encodes the hidden state $h_{i_0}$ of the RNN translator at a given step $i_0$ to the latent variable $g_i$, and then tries to reconstruct the hidden states $h_i$ for subsequent steps ($i\geq i_0$) by evolving and decoding the latent variable. The update equations are given by
\eqs{g_{i_0}&=E(h_{i_0}),\\
g_i&=\tilde{W}(V_i)\cdot g_{i-1},\quad(i=i_0+1,i_0+2,\cdots)\\
h'_i&=D(g_i),\quad(i=i_0,i_0+1,i_0+2,\cdots)}
where $E$ and $D$ represent the encoder and decoder maps respectively. Here the RAE hidden state $g_i\in\dsR^{\tilde{d}}$ is updated by an linear transformation $\tilde{W}(V_i)$ that will still depend on the input potential sequence $V_i$, as illustrated in \figref{fig:RAE}(b). The encoder and the decoder are implemented by feedforward networks as shown in \figref{fig:RAE}(c). The RAE is trained to minimize the reconstruction loss
\eq{\scL_\text{RAE}=\sum_{i\in\text{window}}(h'_i-h_i)^2.}
It is important that the RAE (knowledge distiller) hidden state $g_i$ has a smaller dimension $\tilde{d}$ compared to the dimension $d$ of the RNN (translator) hidden state $h_i$, therefore it can enforce an information bottleneck that only allows the vital information to be passed down in $g_i$. Furthermore, instead of using a single auto-encoder to compress the hidden state at each step independently, the RAE connects a series of decoders together by a recurrent neural network. This design is to ensure that the latent representation $g_i$ remains coherent among a series of steps and contains the key variables that should be passed down along the sequence. A similar RAE architecture was proposed in Ref.\,\onlinecite{Mirowski2010Dy} and recently redesigned in Ref.\,\onlinecite{Iten:2018Di} to enable AI scientific discovery on sequential data. In this way, the RAE compresses the original RNN to a more compact RNN capturing the most essential information and its induced update rules.

\begin{figure}[t]
\begin{center}
\includegraphics[width=0.86\columnwidth]{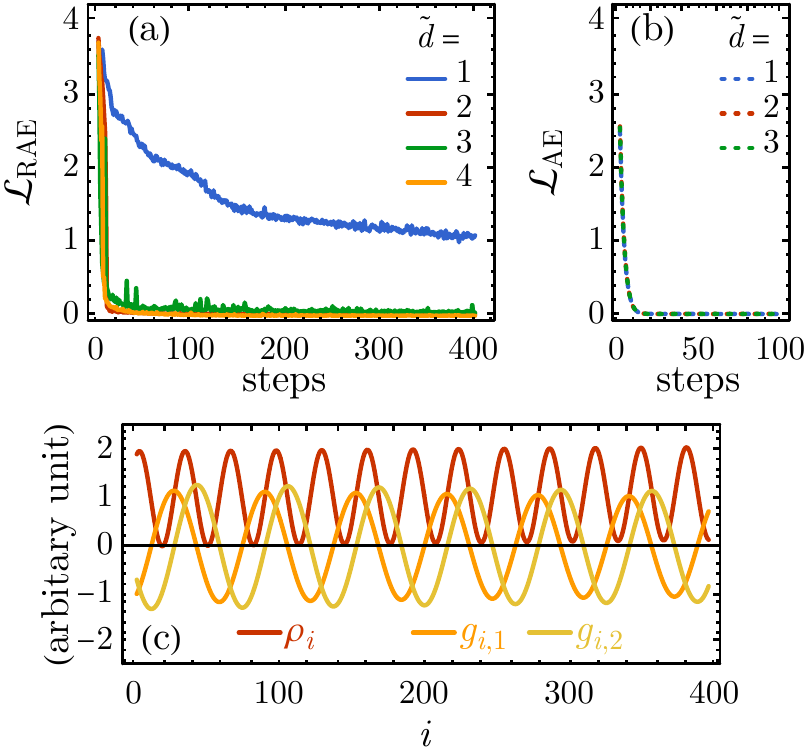}
\caption{(a) The RAE reconstruction loss $\scL_\text{RAE}$ v.s. the training steps for the quantum case. Different curves are for different RAE hidden state dimensions $\tilde{d}$. $\tilde{d}=2$ turns out to be the minimal $\tilde{d}$ without  sacrificing the reconstruction loss. (b) The AE reconstruction loss $\scL_\text{AE}$ v.s. training steps for the classical thermal gas. The vanishing $\scL_\text{AE}$ implies that there is no need to pass any variable along the sequence in this case. (c) The RNN output density profile $\rho_i$ and the RAE hidden state $g_i=(g_{i,1},g_{i,2})$ for a constant potential $V_i=1$. It shows that the periodicity of $g_i$ is twice of $\rho_i$.}
\label{fig:loss}
\end{center}
\end{figure}

As shown in \figref{fig:loss}(a), we find that the reconstruction loss $\scL_\text{RAE}$ of the RAE increases dramatically only when its hidden state dimension $\tilde{d}$ is squeezed below two (i.e. $\tilde{d}<2$), implying that the key feature can be stored in a two-component real vector (i.e. $\tilde{d}=2$) in the most parsimonious manner, as $g_i=(g_{i,1},g_{i,2})$. Here we show that $g_i$ in fact represent the quantum wave function and its first order derivation. The evidences are two fold: 

First, we try to use the trained RNN to predict the density with a constant potential $V$, the result of which should be $\cos^2(kx_i)$ with $k=\sqrt{-V}$ being the momentum. If $g_{i,1}$ and $g_{i,2}$ are the wave function and its derivative, it should be $\cos(kx_i)$ and $\sin(kx_i)$, respectively, whose periods are twice of the period of $\rho_i$ with phases shifted by $\pi/2$ relative to each other. As shown in \figref{fig:loss}(c), $g_i$ indeed displays the periodicity doubling and the relative phase shift. 

Second, we open up the recurrent block of the RAE to extract the update rules for $g_i$, which is machine's formulation of the physical rules. The update rules are encoded in the transformation matrix $\tilde{W}(V_i)=\sum_{n=0}^{n_W}\tilde{W}^{(n)}V_i^n$, which are parameterized by the Taylor expansion coefficient matrices $\tilde{W}^{(n)}$. To connect this formulation to the Schr\"odinger equation we familiar with, we notice that this mapping is invariant under a linear transformation $M\in\mathrm{GL}(2,\dsR)$ applied to all $\tilde{W}^{(n)}$.  We find that it is always possible to find a proper linear transformation that can \emph{simultaneously} bring all $\tilde{W}^{(n)}$ to the following form
\eqs{\label{eq:WRAE}
M^{-1}\tilde{W}^{(0)}M&=\mat{0.9993&0.1007\\0.0013&0.9987}\approx\mat{1&a\\0&1},\\
M^{-1}\tilde{W}^{(1)}M&=\mat{0.0067&0.0004\\0.1001&0.0024}\approx\mat{0&0\\a&0}.}
Here the numerical matrix elements are what we obtained from a particular instance of the trained RAE. They can be associated to the lattice constant $a$ to the leading order given that $a=0.1$, and we have also verified that they scale correctly with $a$ as proposed. The result in \eqnref{eq:WRAE} points to the following difference equation
\eq{\label{eq:Schrodinger}
\mat{g_{i+1,1}\\ g_{i+1,2}}=\mat{1&a\\aV_i&1}\mat{ g_{i,1} \\ g_{i,2}}.}
If we interpret $g_{i,1}$ as the quantum wave function $\psi(x_i)$ and $g_{i,2}$ as its first order derivative $\partial_x\psi (x_i)$, \eqnref{eq:Schrodinger} corresponds to a discrete version of the Schr\"odinger equation $\partial_x^2\psi(x)=V(x)\psi(x)$ as the particle energy was taken to be zero. So the RAE identifies two real numbers as the essential variables in the hidden states. They can be interpreted as the quantum wave function and its first order derivative. Their update rule is consistent with the Schr\"odinger equation.

In this way, without any prior knowledge of quantum mechanics, the introspective learning architecture can develop the concept of the quantum wave function and discover the Schr\"odinger equation when it is only provided with experimental data of potential and density pairs. As a consistency check, we train the same introspective recurrent neural network on the potential and density data of the high-temperature thermal gas following $\rho_i\propto e^{-\beta V_i}$ at a fixed inverse temperature $\beta$. In this case, we can even reduce the RAE to an auto-encoder (AE) without sacrificing the reconstruction loss $\scL_\text{AE}$. As shown in \figref{fig:loss}(b), the $\scL_\text{AE}$ remains vanishing for any $\tilde{d}$, implying that there is no need to pass any variable along the sequence and hence the Sch\"odinger equation will not emerge for thermal gas.

In conclusion, we design the architecture that combines a task machine directly learning the experimental data and an introspective machine working on the neural activations of the task machine. The separation of the task machine from the introspective machine effectively isolates the knowledge distillation from affecting the task performance, such that the whole system can simultaneously improve the task performance and approach the parsimonious limit of knowledge representation, without trading off between one another. Here we show that this architecture can discover the Schr\"odinger equation from the potential-to-density data. Therefore we name it as the ``Schr\"odinger machine''. We envision that the same architecture can be generally applied to other machine learning applications to physics problems and enable machine learning to discover new physics in the future. 

Besides, there are another few points worth highlighting in this work. First, although the use of Taylor expansion for the non-linear functions in our RNN is not essential and can be replaced by neural network models, it has the advantage of being analytical tractability which makes it easier to understand how the RNN works. Second, the potential-to-density mapping is also an essential component in the density functional theory, known as the Kohn-Sham mapping.\cite{Kohn:1965Se} The existing machine learning solutions for this task include the kernel method and the convolutional neural network approach.\cite{Snyder:2012Fi, Li:2014Un, Li:2016Pu, Brockherde:2017By, Khoo:2017So} The RNN approach introduced here has the advantage of being spatially scalable without retraining, which could find potential applications in boosting the density functional calculation and material search. Thirdly, we invent a model that incorporates the auto-encoder with the recurrent neural network, which can find a compact representation of the entire RNN model. This algorithm can find its application in other occasions of model compression and knowledge transfer.  

{\bf Acknowledgement.} This work is funded mostly by Grant No. 2016YFA0301600 and NSFC Grant No. 11734010. CW acknowledges the support of the China Scholarship Council. YZY acknowledges the stimulating discussion with Da Xiao, Lei Ma and Mingli Yuan in the 2017 and 2018 Swarma Club Kaifeng Research Camp.

\bibliography{ML}
\bibliographystyle{apsrev5}

\newpage
\onecolumngrid
\section{Supplemental Material}
\twocolumngrid
\appendix
\section{Data Acquisition}

The data for training RNN  are generated by solving the ``simplified" Schr\"odinger equation in 1d
\begin{equation}
V(x)\psi(x) = \partial_{x}^2\psi(x).
\end{equation}
$x$ labels the position in 1D. The potential begins at $x=0$ and $V(x_i) = V_{i}$ for $x_i\equiv i a$ where $a=0.1$ is a short range cut-off. We define  $k_{i} = \sqrt{-V_{i}}$, then the wave function should take the form of 
$\psi(x) = A_{i} \sin(k_{i}x) + B_{i} \cos(k_{i}x)$ for ${x_i\leq x<x_{i+1}}$. Matching the wave function and its derivative will give the relations,
\begin{equation}
\begin{aligned}
k_{i+1}A_{i+1} = & A_{i}(k_{i+1}\sin(k_{i}x_i)\sin(k_{i+1}x_i) \\
&+ k_{i}\cos(k_{i}x_i)\cos(k_{i+1}x_i))\\
&+B_{i}(k_{i+1}\cos(k_{i}x_i)\sin(k_{i+1}x_i)\\
&-k_{i}\sin(k_{i}x_i)\cos(k_{i+1}x_i))\\
\end{aligned} \label{data1}
\end{equation}
\begin{equation}
\begin{aligned}
k_{i+1}B_{i+1} = & B_{i}(k_{i}\sin(k_{i}x_i)\sin(k_{i+1}x_i) \\
&+ k_{i+1}\cos(k_{i}x_i)\cos(k_{i+1}x_i))\\
&+A_{i}(k_{i+1}\sin(k_{i}x_i)\cos(k_{i+1}x_i)\\
&-k_{i}\cos(k_{i}x_i)\sin(k_{i+1}x_i))\\
\end{aligned}  \label{data2}
\end{equation}
With these relations, we can solve all the $A_{i}, B_{i}$ starting from a fixed initial condition $A_{0} =1,B_{0}=1 $, hence we can construct the wave function $\psi(x)$. Finally the density at $x_i$ is given by
\begin{equation}
\rho_{i}=\psi(x_i)^2.  \label{data3}
\end{equation}
In summary, each data is generated in following steps:\
\begin{enumerate}
\item{Set $V_{1} = -1$ and the rest $V_{i} = -2*\mathsf{rand} -R$. Where $\mathsf{rand}$ is a random number uniformly distributed in $[0,1]$ for each $V_{i}$, and $R$ is a random number uniformly distributed in $[0,1]$ which is the same for each sequence. We use $R$ to randomly shift the energy scale for each data.}\\
\item{Make the potential $V_{i}$ more smooth by performing a flatten operation, $V_{i+1} = 0.5 * (V_{i} + V_{i+1})$, for $q$ times, where $q$ is a random integer between $1$ and $20$.\\}
\item{Get the density sequence $\rho_{i}$ for this potential by solving \eqnref{data1},\eqnref{data2} and using \eqnref{data3}.\\}
\end{enumerate}
In practice, we collect 15000 data, 10000 of them used for training and 5000 of them are used for validation.

While the potential data for RAE are generated in the same way as for RNN, and the hidden state ${h_{i}}$ are collected by evolving the trained RNN.  We collect 15000 data, 10000 of them are used for training and 5000 of them are used for the validation.

\section{Network Parameters}

We elaborate on the details of our training process. For the RNN based on Taylor expansion, we cut off the expansion at power $n_{W} = 2$, and consider the hidden space dimension $d$ from 1 to 6. Taking $d =6$ as an example, the initial $h_0 = (1,1,1,1,1,1)$ and the vector $p= (p_1,0,0,0,0,0)$ without loss of generality, the parameter $p_1$ is set to be $1$ initially.  We initialize the coefficient matrices $W^{(n)}$ to
\eqs{
W^{(0)}&=\mathbf{1}_{d\times d}+\frac{0.01}{d}\mathsf{randn}_{d\times d},\text{ (for $n=0$)}\\
W^{(n)}&=\frac{0.01}{d}\mathsf{randn}_{d\times d},\text{ (for $n>0$)}}
where $\mathbf{1}_{d\times d}$ stands for the $d\times d$ dimensional identity matrix and $\mathsf{randn}_{d\times d}$ stands for the $d\times d$ dimensional random matrix whose elements follow independent Gaussian distributions (with unit variance and zero mean). We use the ADAM optimizer with learning rate $0.0002$. The mini-batch size is $5$. The training window is from $i=5$ to $i=55$.

For the RAE network, the encoder is a feedforward network of $d=6 \to 100 \to \mathsf{ramp} \to  \tilde{d} $ structure and the decoder is also a feedforward network of $\tilde{d} \to 100 \to \mathsf{ramp} \to d=6$ structure. We use the ADAM optimizer with learning rate $0.001$. The mini-batch size is $5$. The training window is from $i=5$ to $i=60$.

\section{Analysis of RNN Translator Loss}

The RNN translator may not be able to formulate physical laws in the most parsimonious language. The hidden state of the RNN may contain redundant information. In fact, there is an analytically tractable limit where we can explicitly demonstrate this possibility. For example, the RNN may tried to capture the differential equation for the density profile directly, instead of that for the quantum wave function. To simplify the analysis, let us take $\hbar^2/(2ma^2)$ as our energy unit and define the potential energy with respect to the single-particle energy level, then the Schr\"odinger equation for the BEC wave function $\psi(x)$ takes a rather simple form of $\partial_x^2\psi(x)=V(x)\psi(x)$. However, in terms of the density profile $\rho(x)=|\psi(x)|^2$, the Schr\"odinger equation implies
\eq{\label{eq:drho}
\partial_x\mat{\rho(x)\\ \eta(x) \\ \xi(x)}=\mat{0 & 2 & 0\\V(x) & 0 &1\\0& 2V(x)&0}\mat{\rho(x)\\ \eta(x) \\ \xi(x)},
}
where $\eta(x)=\Re\psi^*(x)\partial_x\psi(x)$ and $\xi(x)=|\partial_x\psi(x)|^2$ are two other real profiles that combine with $\rho(x)$ to form a system of linear differential equations. The recurrent rule for such a system lies within the description power of our RNN architecture. If the RNN choose to identify its hidden state as $h_i=[\rho(x_i),\eta(x_i),\xi(x_i)]^\intercal$, the following parameters will allow it to model \eqnref{eq:drho} with good accuracy to the first order in $a$:
\eq{\label{eq:Wp_base}
W^{(0)}=\mat{1&2a&0\\0&1&a\\0&0&1}, W^{(1)}=\mat{0&0&0\\a&0&0\\0&2a&0},p=\mat{1\\0\\0}.
}
This theoretical construction at least provides us a base RNN that demonstrates why the proposed architecture could work in principle. The performance can be further improved by relaxing the parameters from this idea limit or by enlarging the hidden state dimension $d$.

\begin{figure}[htbp]
\begin{center}
\includegraphics[width=0.68\columnwidth]{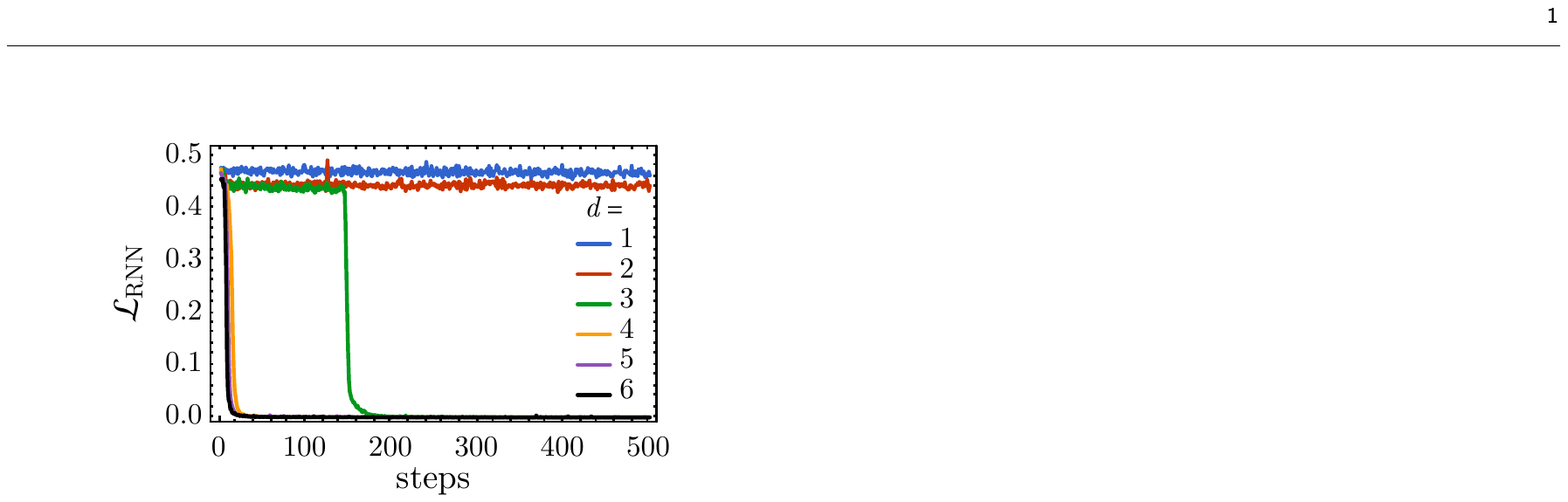}
\caption{The RNN translator loss $\scL_\text{RNN}$ (on the test data set) v.s. the training steps, for different hidden state dimensions $d=1,2,\cdots, 6$. The RNN is only able to master the potential-to-density mapping for $d\geq3$.}
\label{fig:loss RNN}
\end{center}
\end{figure}

However, what is the minimum hidden state dimension $d$ (in terms of real variables) for the RNN to function well in the potential-to-density mapping? Can the RNN discover that the quantum wave function $\psi(x)$ could provide a more parsimonious description, which only requires two real variables $\Re\psi(x)$ and $\Im\psi(x)$ to parameterize? To answer these questions, we train the RNN translator under different hidden state dimensions $d$. As shown in \figref{fig:loss RNN}, we observe that the loss $\scL_\text{RNN}$ only drop significantly if $d\geq 3$, implying that the RNN was unable to realize the more efficient ($d=2$) wave function description. For the $d=3$ case, as we read out the hidden states $h_i$ at each step, we found that they indeed correspond to the vector $[\rho(x_i),\eta(x_i),\xi(x_i)]^\intercal$ up to specific linear transformation (depending on the random initialization of the model parameters), confirming that the RNN indeed works like the base model \eqnref{eq:Wp_base}. From this example, we see that the RNN could develop legitimate and predictive rules of physics, such as \eqnref{eq:drho}, from the observation data. It tends to work directly with the variables present in the observation data to get the job done. Sometimes the rules it found can work well enough that the RNN may not have the motivation to develop higher-level concepts like quantum wave functions.

\end{document}